\documentclass[twocolumn,showpacs,preprintnumbers,amsmath,amssymb,floatfix]{revtex4}
\usepackage{amssymb}
\usepackage{epsfig}

\begin{document}
\title{The Failure of the Ergodic Assumption}
\author{M. Ignaccolo$^{1}$, M. Latka$^{2}$ and B.J. West$^{1,3}$}
\begin{abstract}
The well established procedure of constructing phenomenological ensemble
from a single long time series is investigated. It is determined that a time
series generated by a simple Uhlenbeck-Ornstein Langevin equation is mean
ergodic. However the probability ensemble average yields a variance that is
different from that determined using the phenomenological ensemble (time
average). We conclude that the latter ensemble is often neither stationary
nor ergodic and consequently the probability ensemble averages can
misrepresent the underlying dynamic process.
\end{abstract}

\maketitle

\address{1) Physics Department, Duke University, \\
Durham, NC\\
2) Institute of Biomedical Engineering, \\
Wroclaw University of Technology, Wroclaw, Poland\\
3) Information Science Directorate, Army Research Office\\
Durham, NC 27709}

The ergodic hypothesis, which states the equivalence of time averages and
phase space averages, began with Boltzmann \cite{boltzmann}, who conjectured
that a single trajectory can densely cover a surface of constant energy in
phase space. This 'proof 'of the hypothesis as well as many subsequent
proofs were subsequently shown to be fatally flawed. It was not until metric
decomposability was introduced by Birkoff \cite{birkoff} that a rigorous
mathematical theory of ergodicity began to take shape. Kinchin \cite{kinchin}%
, who wrote a seminal work on the mathematical foundations of statistical
mechanics, offered a second meaning for the ergodic hypothesis, that being,
to assume the truth of the hypothesis and judge the theory constructed on
this basis by its practical success or failure. This latter perspective is
the one adopted by the vast majority of physicists, with the subsequent
replacement of phase space averages with averages over an ensemble
probability density. We call this latter use of ergodicity the 'ergodic
assumption'.

The ergodic assumption was quite evident in quantum measurements, which
historically employed single particle ensembles to evaluate averages.
However the advances in instrumentation and measurement techniques over the
past decade or so enabled the measurement of single molecule time series and
these relatively recent experiments forced the ergodic assumption out of the
shadows into the foreground. Margolin and Barkai \cite{margolin} have argued
that a wide variety of complex phenomena whose dynamics consist of random
switching between two states, such as the fluorescent intermittency of
single molecules \cite{hasse} and nanocrystals \cite{nirmal} may display
non-ergodicity, which is to say the ergodic assumption breaks down. The
break down occurs because the probability density for sojourning within one
of the two states is given by the inverse power law $\psi \left( t\right)
\varpropto t^{-\mu }$ so that when $\mu <2$ the average sojourn time within
a state diverges and the time series is manifestly non-ergodic 
\index{west08}. A similar breakdown is obseved in neuroscience where the
cognitive activity associated with perceptual judgement is determined to
span the interval $1<\mu <3$ \cite{grigolini} and consequently to manifest
non-ergoic behavior in a variety of experiments \cite{gilden,correll}. 

In this Letter we show that the ergodic assumption is even more delicate
than determined in these experiments, which is to say it is less applicable
to complex networks in general, or physical networks in particular, than was
previously believed. To establish the extent of the failure of the ergodic
assumption we begin with a review of its introduction into the physics
literature.

A stochastic physical process is denoted by a dynamic variable \textit{X(t) }%
that can be the velocity of a particle undergoing Brownian motion, the
erratic voltage measured in an electroencephalogram or the time series from
any of a vast number of other complex physical or physiological phenomena.
In any event in the absence of direct evidence to the contrary it is assumed
that the underlying process is stationary in time implying that this
observable can be determined experimentally from a single historical record 
\textit{X(t)} measured over a sufficiently long time. As stated in a classic
paper by Wang and Uhlenbeck (WU) in 1945 \cite{wang}:

\begin{quote}
..One can then cut the record in pieces of length T (where T is long
compared to all "periods" occurring in the process), and one may consider
the different pieces as the different records of an ensemble of
observations. In computing average values one has in general to distinguish
between an ensemble average and a time average. However, for a stationary
process these two ways of averaging will always give the same result...
\end{quote}

This technique for generating empirical realizations of physical ensembles
subsequently proved to be very useful and has been the form of the ergodic
assumption adopted to determine average quantities from time series for
three generations of scientists. The stationarity assumption implies that
single variable distributions are independent of time, so that their
ensemble averages and time averages are the same. Moreover, the joint
distribution functions \ for pairs of dynamic variables depends only on the
difference in time between the measurements of the two variables. Formally,
this implies that the difference variable $Z(t,\tau )=X(t+\tau )-X(t)$ is
only a function of the time difference $\tau $, that is, $Z(t,\tau )=Z(\tau
).$ However a stationary time series need not have stationary increments and
the difference variable can depend on $t$ and $\tau $ separately. Since the
introduction of phenomenological ensembles over a half century ago the
phenomena of interest have become more complex and consequently the WU
approach to evaluating averages is often not adequate. To establish this
inadequacy we go back to the prequel of WU, another classic, and consider a
process first analyzed in detail by Uhlenbeck and Ornstein (UO) \cite
{ornstein} in 1930.

Let us consider the UO stochastic process about which we know everything so
as to test the ergodic assumption. Consider the UO Langevin equation 
\begin{equation}
\frac{dX(t)}{dt}=-\lambda X(t)+F\left( t\right)  \label{langevin1}
\end{equation}
where in a physics context $X(t)$ is the one-dimensional velocity, the
particle mass is set to one, $\lambda $ is the dissipation parameter and $%
F(t)$ is a random force. In other contexts the dynamic variable has been
taken to be the voltage measured in an EEG record \cite{massi}, or any of a
number of other interesting observables. The solution to the UO Langevin
equation, as we show below, actually violates the stationarity assumption
made by Wang and Uhlenbeck in 1945. To establish this violation we calculate
the variance using the phenomenological ensemble and find that it is
different from that obtained using the multi-trajectory ensemble (MTE), that
is, the ensemble obtained by solving the stochastic differential equation (%
\ref{langevin1}) using an ensemble of realizations of the random force. The
significance of this result can not be overestimated since it directly
contradicts a half century of analyses made assuming the ergodicity of the
phenomenological ensembles constructed from the time series.

The formal solution of the OU Langevin equation is, with the initial
condition $X(0)$, 
\begin{equation}
X(t)=e^{-\lambda t}\left[ X(0)+\int\limits_{0}^{t}F(t^{\prime })e^{\lambda
t^{\prime }}dt^{\prime }\right].
\label{langsol}
\end{equation}
A proper interpretation of this solution requires the specification of the
statistics of the random force. If we assume the random force is a
zero-centered delta correlated Gaussian process, as is often done, then the
solution (\ref{langsol}) is generally accepted as a stationary ergodic
process.

\bigskip Recall that an ergodic process is one for which the time average of
an analytic function of the dynamic variable

\begin{equation}
\left\langle g(X)\right\rangle _{T}\equiv \underset{T\rightarrow \infty }{%
\lim }\frac{1}{T}\stackrel{T}{\underset{0}{\int }}g(X(t^{\prime
}))dt^{\prime }  \label{ergodic1}
\end{equation}
and the MTE average

\begin{equation}
\left\langle g(X)\right\rangle _{MTE}\equiv \int g(x)P(x)dx\text{ }
\label{ensemble1}
\end{equation}
are equal

\begin{equation}
\left\langle g(X)\right\rangle _{T}=\left\langle g(X)\right\rangle _{MTE}.
\label{ergodic2}
\end{equation}
If $g(x)=x$, then (\ref{ergodic1}) is a generalization of the law of large
numbers and in the limit of long time lags the time average of the dynamic
variable exists in probability \cite{mccauley}. The process in this case is 
\textit{mean ergodic}, or \textit{first-order ergodic}, which requires that
the autocorrelation function have the appropriate asymptotic behavior, that
being, for $C_{X}(\tau )=\left\langle X(t)X(t+\tau )\right\rangle _{T}$ we
must have $\frac{1}{2T}\stackrel{T}{\underset{0}{\int }}C_{X}(\tau )d\tau
\rightarrow 0$ as $T\rightarrow \infty $ in which case (\ref{ergodic2}) is
true \cite{gnedeko}. If we consider the quadratic variable $%
Z(t)=X(t)X(t+\tau )$ then $Z(t)$ is mean ergodic if $\frac{1}{2T}\stackrel{T%
}{\underset{0}{\int }}C_{Z}(\tau )d\tau \rightarrow 0$ as $T\rightarrow
\infty $, which implies that $X(t)$ is second-order ergodic. The question
remains whether the phenomenological ensemble averages constructed from a
single historical trajectory of long but finite length $T$ satisfies (\ref
{ergodic2}) for any $g(x)$ other than linear as explicitly assumed in WU. It
is probably obvious but it should be said that first-order ergodicity is
important in statistical physics because it guarantees that the
microcanonical ensemble of Gibbs produces the correct statistical average in
phase space for Hamiltonian systems \cite{mccauley}.

To answer the question regarding the validity of (\ref{ergodic2}) for
quantities higher than first-order consider the solution to the UO Langevin
equation, which since the dynamic equation is linear has the same statistics
as the random force. Assuming the random force is a zero-centered delta
correlated Gaussian process, we know that the solution (\ref{langsol}) is
completely determined by the mean and variance. The solution with $X(0)=0$
yields the average values for the dynamic variable $\left\langle
X\right\rangle _{T}=\left\langle X\right\rangle _{MTE}=0$. The
autocorrelation function is $C_{X}(\tau )=\frac{\sigma _{F}^{2}}{2\lambda }%
e^{-\lambda \tau }$ so that the condition $\frac{1}{2T}\stackrel{T}{%
\underset{0}{\int }}C_{X}(\tau )d\tau \rightarrow 0$ as $T\rightarrow
\infty $ is satisfied and consequently the solution to the UO Lagevin
equation is first-order ergodic.

Now let us consider the variance using the MTE averages consisting of an
infinite number of trajectories all starting from $X(0)=0$ (without loss of
generality) yielding 
\begin{equation}
\sigma ^{2}\left( t\right) \equiv \left\langle X^{2}\left( t\right)
\right\rangle _{MTE}-\left\langle X\left( t\right) \right\rangle _{MTE}^{2}=%
\frac{\sigma _{F}^{2}}{2\lambda }\left[ 1-e^{-2\lambda t}\right] .
\label{variance}
\end{equation}
Note that (\ref{variance}) is also the variance obtained using the Gaussian
solution to the Fokker-Planck equation for the ensemble probability
distribution.

Now let us examine the single trajectory case. To create an ensemble of
trajectories consider different portions of the single trajectory. Let $X(t)$
($X(0)$$=$0) be a single trajectory of maximum finite length $T$ and
consider the set of trajectories for $t\in [0,T-\tau ]$ 
\begin{equation}
Z(t,\tau )=X(t+\tau )-X(t).\;\;\;\;\;\;\;  \label{pseudoGibbs}
\end{equation}
Using (\ref{langsol}), we can write $X(t+\tau )$ as 
\begin{equation}
X(t+\tau )=e^{-\lambda \tau }\left[ X(t)+\int\limits_{0}^{\tau
}F(t+t^{\prime })e^{\lambda t^{\prime }}dt^{\prime }\right]  \label{langsol1}
\end{equation}
where the initial condition for the trajectory at $t+\tau $ is the time
series at time $t$. Inserting (\ref{langsol1}) into (\ref{pseudoGibbs}), we
obtain for $t\in [0,T-\tau ]$ 
\begin{equation}
Z(t,\tau )=e^{-\lambda \tau }\int\limits_{0}^{\tau }F(t+t^{\prime
})e^{\lambda t^{\prime }}dt^{\prime }+X(t)\left[ e^{-\lambda \tau }-1\right]
\;\;\;\;\;\;\;  \label{pseudoGibbs2}
\end{equation}
Note that the first term of the \textit{rhs} of (\ref{pseudoGibbs2}) is the
solution of the OU Langevin equation~(\ref{langevin1}) with $t$$=$$\tau $
and $X(0)$$=0$, apart from the fact that in the integrand of~(\ref
{pseudoGibbs2}) only the time in the range $[$$t$,$t$$+$$\tau $$]$ are
considered for the function $F$ (instead of the range ($0$,$\tau )$ as in (%
\ref{langsol})). We now define the solution starting from zero to be 
\begin{equation}
X_{0}(t,\tau )=e^{-\lambda \tau }\int\limits_{0}^{\tau }F(t+t^{\prime
})e^{\lambda t^{\prime }}dt^{\prime }  \label{pseudoGibbs3}
\end{equation}
so that the term $X(t)$ in~(\ref{pseudoGibbs2}) is equivalent to $X_{0}(0,t)$%
, and therefore with $t\in [0,T-\tau ]$ (\ref{pseudoGibbs2}) reduces to 
\begin{equation}
Z(t,\tau )=X_{0}(t,\tau )+X_{0}(0,t)\left[ e^{-\lambda \tau }-1\right]
\;\;\;\;\;\;\;  \label{pseudoGibss4}
\end{equation}
This equation reveals that the statistics of the variable $Z(t,\tau )$
depends on the particular value of $t$ through the second term on the 
\textit{rhs}. In fact the term $X_{0}(0,t)$ has zero mean and variance that
depends on $t$ according to (\ref{variance}). Also notice that the first
term of the above equation depends on $t$ \ but the ''dependence''\ is a
time translation of the variable $F$ in the integrand of ~(\ref{pseudoGibbs3}%
). But the random force $F$ is stationary by assumption and thus $%
X_{0}(t,\tau ) $ is statistically equivalent to $X_{0}(0,\tau )$. The
function $Z(t,\tau ) $ is the sum of two statistically independent Gaussian
variables and therefore is itself a Gaussian variable with $\left\langle
Z(t,\tau )\right\rangle _{T}=\left\langle Z(t,\tau )\right\rangle _{MTE}=0\;$%
and second-moment 
\begin{equation}
\left\langle Z^{2}(t,\tau )\right\rangle _{MTE}=\sigma ^{2}(\tau )+\left[
e^{-\lambda \tau }-1\right] ^{2}\sigma ^{2}(t).  \label{pseudoGibbs5}
\end{equation}
We use this variance to demonstrate that the phenomenological ensemble
created is not equivalent to MTE because the trajectories $Z(t,\tau )$ are
not statistically equivalent to those of $X(t)$. We show that $\sigma
_{straj}^{2},$ the variance calculated with the ensemble $Z(t,\tau ),$ is
different from that calculated using the MTE (\ref{variance}).

To calculate the variance $\sigma _{straj}^{2}(\tau )$ using the
phenomenological ensemble, we simply need to calculate the mean value (among
the trajectories) using (\ref{pseudoGibbs5}). Here the variance is
calculated using the time average since only the time series is assumed to
be available to us 
\begin{equation}
\sigma _{straj}^{2}(\tau )=\frac{1}{T-\tau }\int\limits_{0}^{T-\tau
}dt\left\langle Z^{2}(t,\tau )\right\rangle _{MTE}.  \label{variancetraj}
\end{equation}
Inserting the mean of the phenomenological second moment into the integrand
yields

\begin{equation}
\sigma _{straj}^{2}(\tau )=\frac{1}{T-\tau }\int\limits_{0}^{T-\tau
}dt\left\{ \sigma ^{2}(\tau )+\left[ e^{-\lambda \tau }-1\right] ^{2}\sigma
^{2}(t)\right\}
\end{equation}
and using (\ref{variance}) allows to write after integrating over time

\begin{eqnarray}
\sigma _{straj}^{2}(\tau ) &=&\sigma ^{2}(\tau )+\left[ e^{-\lambda \tau
}-1\right] ^{2}  \nonumber \\
&&\times \left[ \frac{\sigma _{F}^{2}}{2\lambda }\right] \left[ 1+\frac{1}{%
T-\tau }\frac{e^{-2\lambda (T-\tau )}-1}{2\lambda }\right]
\end{eqnarray}
Since $\lambda $$T$$\gg $1 (because $T$ is the length of the record) we can
write 
\begin{equation}
\sigma _{straj}^{2}(\tau )\approx \sigma ^{2}(\tau )+\left[ e^{-\lambda \tau
}-1\right] ^{2}\left[ \frac{\sigma _{F}^{2}}{2\lambda }\right]
\label{variancetraj2}
\end{equation}
and substitution from (\ref{variance}) yields

\begin{equation}
\sigma _{straj}^{2}(\tau )\approx \frac{\sigma _{F}^{2}}{\lambda }%
(1-e^{-\lambda \tau }).  \label{final}
\end{equation}
The last expression shows that the phenomenological ensemble trajectories $%
Z(t,\tau )$ behave as a MTE with an effective dissipation that is half the
dissipation rate in the Langevin equation generating the process, that is,

\begin{equation}
\lambda _{eff}=\lambda /2.  \label{final2}
\end{equation}
Consequently, it takes half as long for the MTE variance to decay as it does
for the phenomenological variance of the time series.

\begin{figure}[]
\includegraphics[angle=-90,width=\linewidth]{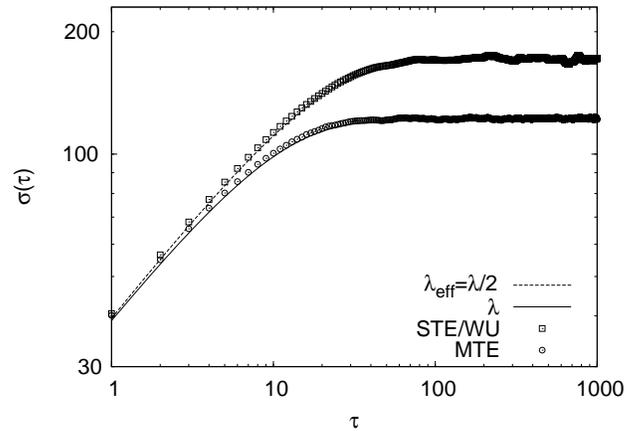}
\caption{The variance calculated for the OU process using the single long
time series and the phenomenological WU ensemble. Note the difference in the
dissipation rates for the two ways of calculating the variance.}
\label{figure1}
\end{figure}

Consider what has been established in this Letter. First a long time series
is generated numerically from the U) Langevin equation. This long series is
separated into a large number of equal length time series as prescribed by
Wang and Uhlenbeck \cite{wang} to form a phenomenological ensemble of
realizations of the UO process. This phenomenological ensemble of
trajectories is shown to be mean ergodic. However when they are used to
calculate the variance of the process anticipating that the same result will
be obtained as that for a MTE average, the probability ensemble average, of
the variance. The expected result turns out to be wrong. The dissipation
parameter determined by the MTE variance is twice that obtained using the
phenomenological ensemble. The inescapable conclusion is that the
phenomenological prescription for calculating the average of any nonlinear
function of the dynamic variable using time series is not equivalent to that
using a MTE. The fact that (\ref{ergodic2}) is not satisfied for $g(x)\neq x$
implies that the ergodic assumption is violated by any finite time
phenomenological ensemble. Given the simplicity of the dynamic model
discussed here; it is mean ergodic; combined with the recent findings
concerning the non-ergodicity of experimental data mentioned earlier, it is
not unreasonable to conclude that the application of the ergodic assumption
in general is probably unwarranted.

\end{document}